\documentclass[12pt]{article}

\usepackage{amsmath}
\usepackage{amsfonts}
\usepackage{amssymb}

\newcommand{\be}{\begin{equation}}
\newcommand{\ee}{\end{equation}}
\newcommand{\ba}{\begin{eqnarray}}
\newcommand{\ea}{\end{eqnarray}}

\begin{document}

\begin{center}
{\large\bf NEW EXACTLY SOLVABLE TWO-DIMENSIONAL QUANTUM MODEL NOT AMENABLE TO SEPARATION OF VARIABLES}\\

\vspace{0.3cm} {\large \bf M.V. Iof\/fe$^{1}$\footnote{e-mail: m.ioffe@pobox.spbu.ru},
D.N. Nishnianidze$^{1,2}$\footnote{e-mail: cutaisi@yahoo.com}, P.A. Valinevich$^{1}$\footnote{e-mail: pavel-valinevich@yandex.ru}
}\\
\vspace{0.2cm}
$^1$Saint-Petersburg State University,
198504 St.-Petersburg, Russia\\
$^2$Akaki Tsereteli State University, 4600 Kutaisi, Republic of Georgia\\
\end{center}
\hspace*{0.5in}
\begin{minipage}{5.0in}
{\small The supersymmetric intertwining relations with second order supercharges
allow to investigate new two-dimensional model which is not amenable to
standard separation of variables. The corresponding potential being the two-dimensional
generalization of well known one-dimensional P\"oschl-Teller model is proven to be exactly solvable
for arbitrary integer value of parameter $p:$ all its bound state energy eigenvalues are found
analytically, and the algorithm for analytical calculation of all wave functions is given.
The shape invariance of the model and its integrability are of essential importance to
obtain these results.
\\
\vspace*{0.1cm} PACS numbers: 03.65.-w, 03.65.Fd, 11.30.Pb }
\end{minipage}

\vspace*{0.4cm}

\setcounter{footnote}{0} \setcounter{equation}{0}
\section{Introduction.}

The beautiful idea of supersymmetry (SUSY) was first introduced \cite{QFT} and developed in
Quantum Field Theory and Elementary
Particle Theory at the seventies of the last century. During these years supersymmetry became one of the most popular and
promising branches of modern High Energy Physics \cite{SUSY-review}.

Supersymmetry was also studied in the simplest toy model of $(0+1)$ Quantum
Field Theory (i.e. in nonrelativistic Quantum Mechanics) in order to clarify some delicate problems of
spontaneous supersymmetry breaking \cite{witten}. Very soon, this by-product of supersymmetrical Quantum
Field Theory became
a new independent tool to study many problems in Quantum Mechanics itself \cite{cooper}. In particular, the notions
of SUSY intertwining relations \cite{cooper}, \cite{intertw} and of shape invariance \cite{shape}
provided both new methods to
derive some old results and to obtain new interesting results. As an example, all previously known
one-dimensional exactly solvable potentials were
reproduced as potentials obeying the shape invariance \cite{dabrowska}. In its turn, SUSY intertwining
relations were successfully
used \cite{david}, \cite{new}, \cite{ioffe}, \cite{poschl} in two-dimensional Quantum Mechanics to obtain a variety
of partially (quasi-exactly) solvable\footnote{By definition,
partial (quasi-exact) solvability of the model means that a part of its energy spectrum and corresponding wave functions
are known. Such models take up an intermediate place between exactly solvable ones and models with unknown
spectra \cite{turbiner}.} models which are not amenable to conventional separation of variables \cite{miller}.
Right up to recent time,
the latter method was the sole practical tool to solve analytically two-dimensional (and higher-dimensional)
quantum problems.

Thus, supersymmetrical approach can be considered as a new method \cite{new}, \cite{ioffe}
to solve (at least,
partially) some two-dimensional Schr\"odinger equations. The procedure can be called
as SUSY separation of variables. It is realizable for the models where equation for zero modes of second order supercharges
allows separation of variables \cite{ioffe}. More of that, after separation of variables one-dimensional equations
must be exactly solvable too. Another procedure of SUSY separation of variables works if one of the partner Hamiltonians
does allow standard separation of variables due to special choice of parameters. Then, SUSY intertwining relations may allow
to obtain eigenfunctions of the second partner Hamiltonian, which is not amenable to standard separation of variables.
This approach was used successfully \cite{PhysRev} for the two-dimensional generalization of Morse potential with
integer or half-integer values of parameter.

The present paper
provides new exactly solvable two-dimensional model with potential depending on three parameters, one of which
has to be integer. To solve the problem, it will be necessary to explore essentially both main ingredients of
SUSY Quantum Mechanics: SUSY intertwining relations and shape invariance. Schematically, to solve the Schr\"odinger
equation with potential $V(\vec x; A,B,p)$ depending on parameters $A,B,p$ three steps will be done. First, to find
such exclusive value of parameter (actually, $p=1,)$ that initial Hamiltonian $H(p=1)$ does allow conventional
separation of variables. Second, using SUSY intertwining relations and shape invariance, to build eigenfunctions
for Hamiltonians $H(p),\,\, p=2,3,...,$ which are not already amenable to separation of variables.
And finally, to prove that all constructed wave functions are normalizable and that no extra levels exist.
The structure of the paper is the following. In Section 2, the model and its main properties are formulated, and
the scheme of investigation is reviewed. The separation of variables for first Hamiltonian $H(p=1)$ is performed
and some delicate properties of potential are discussed in Section 3. The zero modes of the
supercharge are built in Section 4, and they are used for construction of wave functions in Section 5.
Normalizability of wave functions is studied in Section 6, where the absence of any other bound states was
proven. A few examples of wave functions for low values of
parameter $p$ are given in Section 7. Conclusions includes the comparison of obtained results with the limiting case
which is explicitly solvable. Rather cumbersome calculation of coefficients necessary for wave functions
and spectrum are presented in Appendix.

\section{Formulation of the Model and the General Scheme.}

We consider the intertwining relations of the form
\begin{equation}
Q^- H = \widetilde{H} Q^-; \qquad H Q^+ = Q^+ \widetilde{H}, \label{intw}
\end{equation}
where $H$ and $\widetilde{H}$ are two-dimensional Hamiltonians of the
Schr\"odinger type
\begin{equation}
H = -(\partial_1^2 + \partial_2^2) + V(x_1, x_2); \quad \widetilde{H} =
-(\partial_1^2 + \partial_2^2) + \widetilde{V}(x_1, x_2),
\end{equation}
and intertwining operators $Q^\pm$ are second-order differential
operators. A number of models of this kind were investigated in the series
of papers \cite{david}, \cite{new}, \cite{ioffe}, \cite{poschl}. In \cite{PhysRev}, it was proven that one of them -
the generalized two-dimensional Morse -
possesses exact solvability. Here we shall show that one more model \cite{poschl} involved in intertwining
relations (\ref{intw}) is also exactly solvable. It reads:
\begin{multline}
H(p) =-\partial_1^2 - \partial_2^2 - 2p(p - 1)\left(
\cosh^{-2}(x_+) + \cosh^{-2}(x_-) \right) +\\
 k_1\left(
\sinh^{-2}(x_2) - \cosh^{-2}(x_1) \right) + k_2\left(
\cosh^{-2}(x_2) - \sinh^{-2}(x_1) \right),\label{H}
\end{multline}
\begin{multline}
\widetilde{H}(p) =-\partial_1^2 - \partial_2^2 - 2p(p + 1)\left(
\cosh^{-2}(x_+) + \cosh^{-2}(x_-) \right) +\\
 k_1\left(
\sinh^{-2}(x_2) - \cosh^{-2}(x_1) \right) + k_2\left(
\cosh^{-2}(x_2) - \sinh^{-2}(x_1) \right),\label{HH}
\end{multline}
\begin{multline} Q^\pm = \partial_1^2 - \partial_2^2
\pm 2p\left( \tanh(x_+) +
\tanh(x_-) \right)\partial_1 \pm \\
2p\left( \tanh(x_-) - \tanh(x_+) \right)\partial_2 +
4p^2\tanh(x_+)\tanh(x_-) + \\ k_1\left( \sinh^{-2}(x_2) +
\cosh^{-2}(x_1) \right) + k_2\left( \cosh^{-2}(x_2) +
\sinh^{-2}(x_1) \right),\label{kuplus}
\end{multline}
where $p$ and $k_{1,2}$ are real parameters, so far arbitrary, and $x_{\pm}\equiv x_1\pm x_2.$ The Hamiltonians
in (\ref{H}), (\ref{HH}) can be represented
in the form:
\be
H_{P-T}(x_1) + H_{P-T}(x_2) + f(x_1, x_2),
\label{gen}
\ee
where $H_{P-T}(x)$ are well known one-dimensional P\"oschl-Teller Hamiltonians,
and $f(x_1, x_2)$ - specific term mixing $x_1$ and $x_2$ variables in potentials. Due to this
expansion, potentials $V(x_1, x_2),\quad \widetilde V(x_1, x_2)$ may be considered \cite{poschl} as a two-dimensional
generalization of P\"oschl-Teller potential.
These models are shape-invariant \cite{shape} with respect to the parameter $p$:
\begin{equation}
H(p+1) = \widetilde{H}(p). \label{shinv}
\end{equation}
Properties (\ref{intw}) and (\ref{shinv}) will be essential for the proof of
exact solvability of the model for positive integer values of $p$.

We remark that, by construction, these models are integrable since from intertwining
relations (\ref{intw}) it follows that
\begin{equation}
[ H, R ]=0, \quad [ \widetilde{H}, \widetilde{R} ]= 0;\qquad R =
Q^+Q^-,\quad \widetilde{R} = Q^-Q^+, \label{Rop}
\end{equation}
with symmetry operators of fourth order in momenta.

The general scheme
to determine the spectrum of the model (\ref{H}) could be adopted from
the paper \cite{PhysRev}, where the full spectrum of
two-dimensional generalization of Morse potential \cite{new}, \cite{ioffe} was found.
In the present context, the plan of construction could be the following: we start with
$H(p=1)$ and find all normalizable solutions $\Psi(\vec{x}; p=1)$ for the
corresponding Schr\"odinger equation (Section 3) as far as it
is amenable to separation of variables. Then, by means of
intertwining relations (\ref{intw}), we find eigenfunctions
$\widetilde{\Psi}(\vec{x}; p=1)$ of $\widetilde{H}(1).$ In general, they might be of
two types \cite{new}: some of them are inherited from $H(1)$ as:
$\widetilde{\Psi}(\vec{x}; 1) = Q^-(1)\Psi(\vec{x}; 1),$ and others
are zero modes of the intertwining operator:
$Q^+(1)\widetilde{\Psi}(\vec{x}; 1)=0.$ In such a
way we obtain {\it all} eigenfunctions of $H(2).$ Due to the
shape-invariance (\ref{shinv}) of the model, $\Psi(\vec{x}; p+1) =
\widetilde{\Psi}(\vec{x}; p),$ and therefore, we have calculated already
$\Psi(\vec{x}; 2).$ Following this strategy step by step, we expect to find the
eigenfunctions and eigenvalues for the Hamiltonians $\widetilde H(p)=H(p+1)$ with
arbitrary integer values $p = 1,2,... .$ At each step, the full variety of eigenfunctions of $H(p+1)$
will belong to one of two classes: 1)\, each normalizable wave function $\Psi (\vec x; p)$
leads to normalizable wave function
$\Psi (\vec x; p+1)\equiv\widetilde{\Psi}(\vec{x}; p) = Q^-(p)\Psi(\vec{x}; p);$
\, 2)\, the same Hamiltonian $H(p+1)$ has also some number of extra normalizable functions which are
specific linear combinations of zero modes $\Omega (\vec x; p)$ of the operator $Q^+(p).$
We shall see below that this plan has to be modified suitably
for the case of our present model, but the main ideas will be analogous to
that of \cite{new}, \cite{ioffe}, \cite{PhysRev}.

\section{Separation of Variables for $H(p=1)$.}

For the Hamiltonian $H(1)$ the standard procedure of separation
of variables in Cartesian coordinates can be applied. Looking for the
solutions of the Schr\"odinger equation $H(1)\Psi(\vec{x}; 1) =
E\Psi(\vec{x}; 1)$ in the form $\Psi(\vec{x}; 1) =
\eta (x_1)\rho (x_2),$ one obtains two one-dimensional equations for
unknown functions $\rho$, $\eta$
\begin{eqnarray}
-\eta''(x_1) - \left( \frac{k_1}{\cosh^2x_1} + \frac{k_2}{\sinh^2x_1}
\right)\eta(x_1) & = & \varepsilon \eta(x_1);\label{sep1}\\
-\rho''(x_2) + \left( \frac{k_2}{\cosh^2x_2} + \frac{k_1}{\sinh^2x_2}
\right)\rho(x_2) & = & \tilde \varepsilon \rho(x_2),\label{sep2}
\end{eqnarray}
where prime denotes the derivative of the function with respect to
its argument, $\varepsilon + \tilde \varepsilon = E$ is the energy value for $H(1),$ and
both $\varepsilon$ and $\tilde\varepsilon$ must be negative for the discrete part of the spectrum.
Thus, we have to consider solutions of one-dimensional Scr\"odinger equations
(\ref{sep1})-(\ref{sep2}) with P\"oschl-Teller potentials $V_{P-T}(x)$.
It is convenient to replace parameters $k_1, k_2$ by $A, B$ according to
$k_1 \equiv B(B-1);$ $k_2 \equiv -A(A-1) .$ Avoiding the case of fall onto center \cite{landau},
we shall restrict ourselves with reasonably
attracting singularity in $V_{P-T}(x_1), V_{P-T}(x_2)$ with coefficients
$k_1\in (-1/4, 0),\,\, k_2\in (0, 1/4),$ i.e. it is sufficient to take $A,B \in (0,1/2).$
The substitution $\eta(x_1) = \sinh^A(x_1)\cosh^B(x_1) F(x_1),$ and
the subsequent change of variable $x_1$ to $z \equiv -\sinh^2(x_1),$ turns (\ref{sep1})
into the hypergeometric equation for the function $F(z):$
$$
z(1-z)\frac{d^2F(z)}{dz^2} + \left( A+\frac{1}{2} -(A+B+1)z
\right)\frac{dF(z)}{dz} + \left( -\frac{1}{4}\left( A+B \right)^2 -
\frac{1}{4}\varepsilon \right)F(z) = 0 .
$$
The pair of independent solutions for
the given value of $\varepsilon $ reads (see 2.3.1(1) in \cite{beitman}):
\be
\eta^{(1)}_{\varepsilon}(x)=\sinh^A(x) \cosh^B(x)\, _2F_1\left(
\frac{A+B+\sqrt{-\varepsilon}}{2}, \frac{A+B-\sqrt{-\varepsilon}}{2};
A+\frac{1}{2}; -\sinh^2(x)\right);   \label{eta1}
\ee
\be
\eta^{(2)}_{\varepsilon}(x)=\sinh^{1-A}(x) \cosh^B(x)\, _2F_1\left(
\frac{1-A+B+\sqrt{-\varepsilon}}{2}, \frac{1-A+B-\sqrt{-\varepsilon}}{2};
\frac{3}{2}-A; -\sinh^2(x)\right) \label{eta2}.
\ee
The similarity of expressions (\ref{eta1}) and (\ref{eta2}) reflects
the obvious symmetry of potential in (\ref{sep1}) under $A \to (1-A).$
The potential under consideration obeys also the similar symmetry under $B \to (1-B).$ But
the corresponding independent solutions are related to solutions (\ref{eta1}), (\ref{eta2})
according to relations between hypergeometric functions (see 2.1.4(23) in \cite{beitman}).
The formulae analogous to (\ref{eta1}) and (\ref{eta2}) hold also for $\rho_{\tilde \varepsilon}^{(1),(2)}(x_2),$ but
with the necessary changes $A \to B,$ $B \to A$ and $\varepsilon \to \tilde \varepsilon.$

To provide the normalizability of $\Psi(\vec{x};1),$ both $\rho$ and
$\eta$ must be normalizable. To analyze the possible bound states of $H(p=1)$
it will be sufficient to consider the asymptotic behaviour for large $|x_1|, |x_2|.$

The general solution $\eta_{\varepsilon}(x)$ is a linear combination:
\begin{equation}\label{combin}
\eta_{\varepsilon}(x)=\alpha_1\eta^{(1)}_{\varepsilon}(x) + \alpha_2\eta^{(2)}_{\varepsilon}(x)
\end{equation}
with arbitrary constants $\alpha_1, \alpha_2.$ The asymptotic behaviour of the
analytic continuation of
hypergeometric functions for large $z=-\sinh^2x$ (see 2.10(2), 2.10(5) in \cite{beitman})
reads:
\begin{equation}\label{asymp}
_2F_1(a,b;c;z)=B_1(a,b,c)\biggl[ (-z)^{-a} + O((-z)^{-a-1})\biggr] + B_2(a,b,c)\biggl[ (-z)^{-b} +
O((-z)^{-b-1})\biggr],
\end{equation}
where constants $B_{1,2}$ are expressed in terms of Gamma functions:
$$B_1(a,b,c)=\frac{\Gamma (c)\Gamma (b-a)}{\Gamma (b)\Gamma (c-a)};\quad
B_2(a,b,c)=\frac{\Gamma (c)\Gamma (a-b)}{\Gamma (a)\Gamma (c-b)}.$$
Substitution of (\ref{asymp}) into (\ref{eta1}), (\ref{eta2}) and (\ref{combin})
gives asymptotically two groups of terms in $\eta_{\varepsilon}(x) :$ proportional to
$(-z)^{-\sqrt{-\varepsilon}/2}(1+O(z^{-1}))$ and
proportional to $(-z)^{+\sqrt{-{\varepsilon}}/2}(1+O(z^{-1})),$ correspondingly. In order to forbid the growing term
in wave function, one has to require the coefficient to vanish:
\ba
&&
\alpha_1B_2\biggl( \frac{A+B+\sqrt{-\varepsilon}}{2}, \frac{A+B-\sqrt{-\varepsilon}}{2}, A+\frac{1}{2}\biggr)
+\nonumber\\
&&
+\alpha_2B_2\biggl( \frac{1-A+B+\sqrt{-\varepsilon}}{2}, \frac{1-A+B-\sqrt{-\varepsilon}}{2}, \frac{3}{2}-A\biggr) = 0 .
\nonumber
\ea
In general, there are two options to fulfil this requirement:
\ba
&&\alpha_2=0;\quad B_2\biggl( \frac{A+B+\sqrt{-\varepsilon}}{2}, \frac{A+B-\sqrt{-\varepsilon}}{2}, A+\frac{1}{2}\biggr) = 0;
\nonumber\\
&&\alpha_1=0;\quad B_2\biggl( \frac{1-A+B+\sqrt{-\varepsilon}}{2}, \frac{1-A+B-\sqrt{-\varepsilon}}{2}, \frac{3}{2}-A \biggr) = 0;
\nonumber
\ea
In the case of arbitrary $A, B,$
these conditions can be achieved by means of suitable choices of energy values $\varepsilon ,$ due to Gamma functions in
denominators of coefficient $B_2:$ arguments $a$ or $(c-b)$ of these Gamma functions must be equal $-n$ with
$n=0,1... .$ Just this condition might give the energies of bound states.
But one can check easily, that in our present case of parameters $A, B \in (0, 1/2),$
these conditions can not be fulfilled for positive values of $\sqrt{-\varepsilon}$.

Thus, we are not able to kill the growing terms
in asymptotic of $\eta_{\varepsilon}$ and $\rho_{\tilde \varepsilon}.$
Therefore, the Hamiltonian $H(p=1)$ has no bound states because
of asymptotic behaviour at large $|x|,$ and the first source for construction of eigenfunctions of $\widetilde H(1):
\,\, \widetilde\Psi (\vec x; 1)=Q^-(1)\Psi (\vec x;1)$ - does not work.
We stress that this statement depends crucially on a chosen region for values of $A, B,$ which in its turn was dictated
by conditions on both potentials $V_1, V_2,$ simultaneously. Taken separately, these Hamiltonians would have
bound states, but for different values of $A, B.$

We notice also that the conclusion above does not depend on the
behaviour of solutions at the singular point $x=0.$ Nevertheless, we will discuss the $x\to 0$ asymptotic of
$\eta^{(1),(2)}_{\varepsilon}$ here, since
it will be necessary for the analysis in subsequent Sections. The point is that (in contrast to standard
situation of nonsingular potentials) both solutions (\ref{eta1}),
(\ref{eta2}) have zero limit at the origin $x\to 0$ for $A \in (0,1/2).$
Namely, their behaviour is $\eta^{(1)} \sim x^A $ and $\eta^{(2)} \sim x^{1-A}.$ This is the typical situation
of the so called "limit circle" kind 
%"limit circle point" 
(see \cite{rid}, Appendix to Section 10.1),
which was widely discussed in the literature in the context of one-dimensional potential
(the so-called Calogero potential)
$g/x^2$ on the semi-axis or on the whole axis (e.g., see \cite{tsutsui}, \cite{gitman1},
\cite{gitman2}).
In such a case, there is continuous freedom in choosing (among many opportunities)
some "good kind of behaviour" for wave functions. The resulting spectrum of the model depends on this choice,
thereby defining the kind of its quantization. The preferences are usually motivated by physical
arguments \cite{gitman1}, \cite{tsutsui}, \cite{lath}. In the case of Hamiltonian $H(p=1)$ this problem is of
little importance due to asymptotic behaviour of solutions at infinity discussed above. One more remark concerns
the extension of solutions to negative semiaxis: it is reasonable to choose an odd way, taking $\eta (-|x|)=-\eta (|x|).$
This choice provides continuity of the derivative $\eta^{\prime}(x)$ at the origin.

We have to remark that from mathematical point of view, both one-dimensional Schr\"odinger
operator with P\"oschl-Teller potential in (\ref{sep1}), (\ref{sep2}) and two-dimensional operator in
(\ref{H}), (\ref{HH}) produce rather nontrivial
problem. It is possible to check that both of them (in two-dimensional case, due to Green's identity of vector calculus),
are symmetric operators, but in strictly mathematical approach, they are unbounded and not self-adjoint for the
conventional choice of smooth functions with a compact support (dense in $L_2$)
as a domain $D(H_{P-T}(x)).$ Similarly to the analysis
given in \cite{gitman1} for the case $V=\alpha / x^2,$ the self-adjoint extension of $H_{P-T}(x)$ includes
also the functions from $L_2$ with specific asymptotic at the singular point $x=0.$
For details, we refer readers to the papers \cite{gitman1}, \cite{tsutsui},
\cite{gitman2} and references therein, where one will find also the description of some paradoxes induced by too naive
approach to singular potentials of $\alpha / x^2$ type.

\section{Construction of Zero Modes of $Q^+$.}

As it was mentioned above, the second possible source of eigenfunctions for
$\widetilde{H}(p)$ are the zero modes of operator $Q^+(p)$. It
is well known \cite{new}, \cite{ioffe}, that the subspace of zero modes of $Q^+(p)$
is invariant under the action of $\widetilde{H}(p).$ This means that
if $\widetilde{\Omega}_k(\vec{x}; p)$ is the zero-mode of $Q^+(p),$ i.e.
$Q^+(p)\widetilde{\Omega}_k(\vec{x}; p) = 0,$ then due to intertwining
relations (\ref{intw})
\begin{equation}
\widetilde{H}(p)\widetilde{\Omega}_k(\vec{x}; p) = \sum\limits_{i=0}^N
C_{ki} \widetilde{\Omega}_i(\vec{x}; p)\label{Cik}
\end{equation}
($C_{ki}$ is the $(N+1)\times (N+1)$ matrix with complex elements). If
the matrix $C_{ki}$ can be diagonalized by some matrix $B:$
\be
BC=\Lambda B;\quad \Lambda = diag (E_0, E_1, ..., E_N),
\label{B}
\ee
the functions
\begin{equation}
\widetilde{\Psi}_i(\vec{x};p) = \sum_{k=0}^N B_{ik} \widetilde\Omega_k(\vec{x}; p)
\label{Psi}
\end{equation}
are the eigenfunctions of $\widetilde{H}(p):$ $
\widetilde{H}(p)\widetilde{\Psi}_i(\vec{x}; p) = E_i
\widetilde{\Psi}_i(\vec{x}; p).$

At first, one needs to calculate $\widetilde{\Omega}_i(\vec{x}; p).$
For this purpose, it is useful to perform the similarity
transformation, which will help to separate variables:
\begin{equation}
q^+(p) = e^{p\chi(\vec{x})} Q^+(p) e^{-p\chi(\vec{x})};\quad
\widetilde{\Omega}_i =
e^{-p\chi(\vec{x})}\widetilde{\omega}_i,\label{gauge}
\end{equation}
where
$$\chi = \ln \left( \cosh(x_+) \cosh(x_-) \right).$$
After that, the problem $Q^+(p)\widetilde{\Omega}_i(\vec{x}; p)=0$
becomes
\begin{equation}
q^+(p) \widetilde{\omega}_i(\vec{x}; p) = 0,\label{q+}
\end{equation}
where $q^+$ reads
\begin{equation}
q^+ = \partial_1^2 - \partial_2^2 + k_1\left( \sinh^{-2}(x_2) +
\cosh^{-2}(x_1) \right) + k_2\left( \cosh^{-2}(x_2) +
\sinh^{-2}(x_1) \right). \label{q+q+}
\end{equation}

The choice of the function $\chi$ provides that (\ref{q+}) is
amenable to separation of variables in Cartesian coordinates.
The two-dimensional equation (\ref{q+}) is equivalent to the pair
of one-dimensional ones if one takes $\widetilde{\omega}_i =
\eta(x_1)\rho(x_2)$, and it appears that they are exactly the
equations (\ref{sep1})-(\ref{sep2}), but with $\varepsilon = \tilde \varepsilon.$
The solutions can be written as linear combinations of
\begin{equation}
\widetilde{\omega}_{\varepsilon} = \eta_{\varepsilon}(x_1)\rho_{\varepsilon}(x_2),\label{lambdan}
\end{equation}
where $\eta_{\varepsilon}$ (and analogously, $\rho_{\varepsilon}$)
must be built from solutions (\ref{eta1}), (\ref{eta2}).
Of course, we are interested only in
normalizable zero modes $\Omega (\vec x; p)$ of the two-dimensional operator $Q^+(p),$ but
the normalizability condition, in comparison with Section 3, is essentially
less restrictive now:
\ba
&&\int |\widetilde{\Omega}(\vec{x}; p)|^2 d^2x =
\int e^{-2p\chi(\vec{x})} |\eta(x_1)|^2|\rho(x_2)|^2 d^2x =\nonumber \\
&&=\int \left( cosh(x_+) cosh(x_-) \right)^{-2p} |\eta(x_1)|^2|\rho(x_2)|^2 d^2x < \infty. \label{Omeganorm}
\ea
The factor $\exp{(-2p\chi (\vec x) )}$ in (\ref{Omeganorm}) is exponentially decreasing
at infinity in all directions on the plane, and it is able to compensate even growing
functions $\widetilde\omega_{\varepsilon}$. Due to asymptotic equivalence $\cosh x \sim \sinh x$ at infinity,
asymptotical behaviour of the integrand of (\ref{Omeganorm}) can be represented as
$$|\widetilde{\Omega}(\vec{x}; p)|^2 \sim (\cosh x_+\cosh x_-)^{-2p}(\cosh x_+-\cosh x_-)^{2\sqrt{-\varepsilon}}.$$
Therefore, the functions $\widetilde\Omega$ are normalizable for arbitrary
values of $p$ and $\varepsilon ,$ satisfying: $\varepsilon > -p^2.$ This fact has to be taken into account
in calculation of the spectrum of $H(p+1)$ (see Section 5).

But at first, we must define the variety of functions $\widetilde\Omega_{\varepsilon},$ which may be used
for construction of actual wave functions. In this context, functions $\eta_{\varepsilon}(x_1)$ and $\rho_{\varepsilon}(x_2)$
are the auxiliary objects for construction of zero modes $\widetilde\Omega$ according to (\ref{gauge}), (\ref{lambdan}).
Therefore, all four possible combinations can be used, in general. The first of them $\widetilde\Omega^{(1)}_{\varepsilon}(\vec x)=
\exp{\biggl(-p\chi(\vec{x})\biggr)}\eta^{(1)}_{\varepsilon}(x_1)\rho^{(1)}_{\varepsilon}(x_2)$ is:
\ba
&&\widetilde\Omega^{(1)}_{\varepsilon}(\vec x) =
=\pm\left( \cosh(x_+) \cosh(x_-) \right)^{-p}
\sinh^{A}(|x_1|) \cosh^B(|x_1|)\cdot\nonumber \\
&& \cdot\sinh^{B}(|x_2|) \cosh^A(|x_2|)\,_2F_1\left(a_{\varepsilon},
b_{\varepsilon}; A+\frac{1}{2}; z_1\right)\cdot
_2F_1\left(a_{\varepsilon}, b_{\varepsilon}; B+\frac{1}{2}; z_2\right)
\label{first}
\ea
where
\be
a_{\varepsilon}\equiv \frac{A+B-\sqrt{-\varepsilon}}{2};\,\, b_{\varepsilon}\equiv \frac{A+B+\sqrt{-\varepsilon}}{2};\,\,
z_1\equiv -\sinh^2x_1;\,\,
z_2\equiv -\sinh^2x_2, \label{def}
\ee
and the sign $\pm$ depends on a quarter on a plane $(x_1, x_2),$
according to the choice at the end of Section 3. Other zero modes $\widetilde\Omega^{(2)}_{\varepsilon}(\vec x),\,
\widetilde\Omega^{(3)}_{\varepsilon}(\vec x),\,\widetilde\Omega^{(4)}_{\varepsilon}(\vec x)$ are obtained from $\widetilde\Omega^{(1)}_{\varepsilon}(\vec x)$ by means of substitutions of pairs of parameters $\biggl( 1-A, B \biggr)$ for $\widetilde\Omega^{(2)}_{\varepsilon}(\vec x),$ of $\biggl( A, 1-B \biggr)$ for $\widetilde\Omega^{(3)}_{\varepsilon}(\vec x),$
and $\biggl( 1-A, 1-B \biggr)$ for $\widetilde\Omega^{(4)}_{\varepsilon}(\vec x),$ instead of $\biggl( A, B \biggr)$ in
$\widetilde\Omega^{(1)}_{\varepsilon}(\vec x).$

\section{Eigenfunctions of the Hamiltonian $H$.}

In this Section we shall look for linear combinations of zero modes
of $Q^+(p),$ which are simultaneously the eigenfunctions of the Hamiltonian $\widetilde H(p)=H(p+1)$ in (\ref{HH}).
Being interested in the discrete energy spectrum $E_n$, we suppose that the corresponding wave functions
are built from the finite number of zero modes $\widetilde\Omega^{(\gamma)}_{\varepsilon_k}(\vec x; p);\, \gamma =1, 2, 3, 4,$ with parameters
$a_k\equiv a_{\varepsilon_k}$ in (\ref{def}),
numbering by discrete values $k=0,1,2,...,N^{(\gamma )},$ so that the constants $a_k$ are ordered as: $a_0>a_1>...>a_{N^{(\gamma )}} .$
We suppose also that four kinds of such wave functions exist: each is built from the corresponding zero modes
$\widetilde\Omega^{(\gamma )}$ with fixed value of $\gamma $ (the value of $\gamma $ defines behaviour at the origin).
According to (\ref{Cik}),
\be
 \tilde{H}(p)\widetilde\Omega^{(\gamma )}_{\varepsilon_k}(\vec x)=\sum_{i=0}^N C^{(\gamma )}_{ki}
 \widetilde\Omega^{(\gamma )}_{\varepsilon_i}(\vec x),\label{D26}
\ee
where $C^{(\gamma )}_{ki}$ are constants, and $N$ also depend on $\gamma =1,2,3,4.$

Performing with $\widetilde H(p)$ the similarity transformation
analogous to that with $Q^+(p)$ in (\ref{gauge}), one obtains:
$$ \tilde{h}(p)\equiv e^{p\chi(\vec{x})}\widetilde{H}(p)
 e^{-p\chi(\vec{x})}=-\partial_1^2-\partial_2^2 + \hat D -
 \frac{k_1}{\cosh^2x_1}-\frac{k_2}{\sinh^2x_1}
 +\frac{k_2}{\cosh^2x_2} + \frac{k_1}{\sinh^2x_2}-4p^2,$$
where the mixing operator $\hat D$ is defined as:
\be
 \hat D\equiv \frac{2p}{\cosh x_+\cosh x_-}
 (\sinh(2x_1)\partial_1+\sinh(2x_2)\partial_2).\label{TT4}
\ee
Then, exploring (\ref{sep1}), (\ref{sep2}), the action of $\tilde h(p)$ on $\widetilde\omega^{(1)}_{\varepsilon_k}(p)$
(see its definition in (\ref{gauge}) and (\ref{lambdan}))
can be expressed as:
\ba
 &&\tilde{h}(p)\widetilde{\omega}^{(1)}_{\varepsilon_k}(p)=
 2\biggl(2p(A+B)+\varepsilon_k-2p^2\biggr)\widetilde{\omega}^{(1)}_{\varepsilon_k}+
 \sinh^{A}(x_1)\cosh^{B}(x_1)\sinh^{B}(x_2)\cdot \nonumber\\
&& \cdot\cosh^{A}(x_2) \hat D \biggl(\,_2F_1(a_k,b_k;A+1/2;-\sinh^2x_1)\,_2F_1(a_k,b_k;B+1/2;-\sinh^2x_2)\biggr),
\nonumber
\ea
and (\ref{D26}) takes the form:
\ba
 &&2\biggl( 2p(A+B)+\varepsilon_k-2p^2\biggr)\, _2F_1(a_k,b_k;A+1/2;z_1)\, _2F_1(a_k,b_k;B+1/2;z_2)+
 \nonumber\\
 &&+\hat D \biggl(\, _2F_1(a_k,b_k;A+1/2;z_1)\, _2F_1(a_k,b_k;B+1/2;z_2)\biggr)=\nonumber\\
 &&=\sum_{i=0}^N C^{(1)}_{ki}\, _2F_1(a_i,b_i;A+1/2;z_1)\, _2F_1(a_i,b_i;B+1/2;z_2).
\nonumber
\ea

After straightforward calculations, (\ref{D26}) can be rewritten as:
\ba
 &2\biggl(2p(A+B)+\varepsilon_k-2p^2\biggr)\, _2F_1(a_k,b_k;c_1;z_1)\cdot\, _2F_1(a_k,b_k;c_2;z_2)+\nonumber\\&
 +\frac{8a_k b_k p}{1-z_1-z_2}\biggl(\frac{z_1(1-z_1)}{c_1}\, _2F_1(a_k+1,b_k+1;c_1+1;z_1)\cdot\,
  _2F_1(a_k,b_k;c_2;z_2)+
 \nonumber\\&
 +\frac{z_2(1-z_2)}{c_2}\cdot\, _2F_1(a_k+1,b_k+1;c_2+1;z_2)\cdot\, _2F_1(a_k,b_k;c_1;z_1)\biggr)=\nonumber\\&
 =\sum_{i=0}^N C^{(1)}_{ki}\cdot\, _2F_1(a_i,b_i;c_1;z_1)\,\cdot\, _2F_1(a_i,b_i;c_2;z_2), \label{6}
 \ea
 where
\be
c_1=A+\frac{1}{2}, \, c_2=B+\frac{1}{2}, \quad
 b_k\equiv b_{\varepsilon_k} = \frac{A+B+\sqrt{-\varepsilon_k}}{2}.
\label{666}
\ee
For $z_2=0$ (\ref{6}) reads:
 \ba
 &2\biggl(2p(A+B)-2p^2+\varepsilon_k\biggr)\, _2F_1(a_k,b_k;c_1;z_1)+
 \frac{8a_k b_k p}{c_1}z_1\cdot\, _2F_1(a_k+1,b_k+1;c_1+1;z_1)=\nonumber\\&
 =\sum_{i=0}^N C^{(1)}_{ki}\cdot\, _2F_1(a_i,b_i;c_1;z_1).\label{7}
 \ea
In the $z_1\to-\infty$ limit, the largest power in the l.h.s. of
(\ref{7}) is $(-z_1)^{-a_k}.$ Therefore,
 \ba
 C^{(1)}_{k,i}=0 \,\, for\,\, i>k, \label{8}
 \ea
i.e. Eq.(\ref{7}) takes the form:
 \ba
 &2 \biggl( 2p(A+B)-2p^2 + \varepsilon_k \biggr)\cdot\, _2F_1(a_k,b_k;c_1;z_1)+
 \frac{8a_k b_k p}{c_1}z_1\,\cdot\, _2F_1(a_k+1,b_k+1;c_1+1;z_1)=\nonumber\\&
 =\sum_{i=0}^kC^{(1)}_{ki}\,\cdot\, _2F_1(a_i,b_i;c_1;z_1).\label{9}
 \ea
Further, for particular values $z_1=0$ and $k=0,$ it gives:
 \be
 C^{(1)}_{00}=2\biggl( 2p(A+B)-2p^2+\varepsilon_0 \biggr).\label{10}
 \ee
Substitution of (\ref{10}) back into (\ref{9}) with arbitrary $z_1$ and $k=0$
leads to $a_0\cdot b_0=0.$ Since all $b_k$ are positive, the only opportunity is
$a_0=0.$
Comparing next powers in Eq.(\ref{9}) for $z_1\to -\infty ,$ we obtain:
$$ a_k+1=a_{k-1}.$$
Together with $a_0=0,$ this relation uniquely defines all values of $a_k:$
\be
a_k=-k;\,\, k=0,1,...,N^{(1)}. \label{ak}
\ee

In turn, comparison of coefficients of $(-z_1)^{-a_k}$ in Eq.(\ref{9}), gives values of elements $C^{(1)}_{kk}.$
Due to (\ref{8}), matrix $C^{(1)}_{ki}$ is triangular. Its diagonal elements $C^{(1)}_{kk}$ coincide
with elements of diagonal matrix $\Lambda $ in (\ref{B}), and therefore, $C^{(1)}_{kk}$ gives a part of the eigenvalues
of discrete energy spectrum of the Hamiltonians $\widetilde{H}(p)=H(p+1):$
 \be
 \widetilde E^{(1)}_k(p)=E^{(1)}_k(p+1)=-2 \biggl( (A+B+2k-p)^2+p^2 \biggr); \,\, k=0,1, ..., N^{(1)}.\label{energy}
 \ee

It is clear from (\ref{energy}) that the lowest energy state corresponds to the maximal
$k,$ i.e. to $k=N^{(1)},$ which can be defined from conditions of normalizability of $\widetilde\Omega^{(1)}_{e_k}(\vec x).$
These conditions were formulated in Section 4, and they can be rewritten now as:
\begin{equation}\label{condition}
  k^{(1)} < \frac{1}{2}(p-A-B);\quad N^{(1)}=\biggl[\frac{1}{2}(p-A-B)\biggr],
\end{equation}
where $[c]$ means the integer part of $c.$

Analogously one can construct three other kinds of energy levels $\widetilde{E}_k^{(\gamma )}(p)$ of
$\widetilde{H}(p)=H(p+1)$ by replacing everywhere above $(A, B)$ by $(1-A, B),$ or by $(A, 1-B),$ or by $(1-A, 1-B),$
correspondingly. The result is the following:
\ba
\widetilde E^{(2)}_k(p)&=&E^{(2)}_k(p+1)=-2 \biggl( (1-A+B+2k-p)^2+p^2 \biggr); \,\, k^{(2)}=0,1, ..., N^{(2)};\label{energy2}\\
\widetilde E^{(3)}_k(p)&=&E^{(3)}_k(p+1)=-2 \biggl( (1+A-B+2k-p)^2+p^2 \biggr); \,\, k^{(3)}=0,1, ..., N^{(3)};\label{energy3}\\
\widetilde E^{(4)}_k(p)&=&E^{(4)}_k(p+1)=-2 \biggl( (2-A-B+2k-p)^2+p^2 \biggr); \,\, k^{(4)}=0,1, ..., N^{(4)},\label{energy4}
\ea
where:
\ba
k^{(2)} &<& \frac{1}{2}(p-1+A-B); \label{condition2} \\
k^{(3)} &<& \frac{1}{2}(p-1-A+B); \label{condition3} \\
k^{(4)} &<& \frac{1}{2}(p-2+A+B). \label{condition4}
\ea
The energy spectra and the corresponding wave functions for several lowest values of $p$ will be given in Section 7.

According to (\ref{Psi}), the eigenfunctions of $H(p+1)$:
\be
 \Psi^{(\gamma )}_k(\vec x; p+1)=\widetilde{\Psi}^{(\gamma )}_{k}(\vec{x}; p)=\sum_{i=0}^N
 B^{(\gamma )}_{ki}\widetilde\Omega^{(\gamma )}_i(\vec{x}; p);\,\, \gamma =1,2,3,4
\label{D37}
\ee
can be written explicitly only after calculation of coefficients $C^{(\gamma )}_{ki}$ in (\ref{D26}) and, after that, of
$B^{(\gamma )}_{ki}$ from (\ref{B}). The first step - calculation of $C^{(\gamma )}_{ki}$ with general $k,i$ - is given in Appendix,
while the calculation of $B^{(\gamma )}_{ki}$ seems to
be rather complicated in a general form. Instead, this will be done explicitly for small values of $p$ in Section 7.

Thus, each Hamiltonian $H(p+1)$ (for $N^{(\gamma )}\geq 0$) possesses $N^{(1)}+N^{(2)}+N^{(3)}+N^{(4)}+4$
bound states $\Psi^{(\gamma )}_k(\vec x; p+1)$ with
energy levels $E^{(\gamma )}_k(p+1),\,\, k=0,1,...,N^{(\gamma )},\,\, \gamma =1,2,3,4,$ which are absent in the spectrum of $H(p).$
As we know, due to SUSY intertwining relations (\ref{intw}) (see also Section 2),
each of these wave functions produce the tower of extra eigenfunctions for higher
Hamiltonians $H(p+n+1),\,\,n=1,2,...$ with the same energy values $E^{(\gamma )}_k(p+1)$. These wave functions
$\Psi^{(\gamma )}_{kn}(\vec x; p+n+1)$ are built by the action of $n$ operators $Q^-(p+m),\, m=1,2,...,n :$
\be
 \Psi^{(\gamma )}_{kn}(\vec x; p+n+1)=\widetilde\Psi^{(\gamma )}_{kn}(\vec{x};p+n)=Q^-(p+n)Q^-(p+n-1)...Q^-(p+1)\Psi^{(\gamma )}_k(\vec{x};p+1),
\label{D38}
\ee
and their indices indicate the number $k$ among $N^{(\gamma )}$ bound states of original Hamiltonian $H(p+1),$ and
the number $n$ of one after another acting operators $Q^-.$

\section{Normalizability of the wave functions.}

According to results of previous Sections, the Hamiltonian $H(p+1)=\widetilde H(p)$
has two classes of bound state wave functions. The second one (in terminology of Section 2)
$\Psi^{(\gamma )}_k(\vec x; p+1),\,k=0,1,...,N^{(\gamma )}$
is produced by normalizable zero modes of $Q^+$ via their suitable linear combinations. The
first class is obtained from the eigenfunctions of lower Hamiltonians by means of operators $Q^-.$
In notations introduced above, they are:
\ba
 \Psi^{(\gamma )}_{m,(p-n)}(\vec{x};p+1)&=&Q^-(p)Q^-(p-1)...Q^-(n+1)\Psi^{(\gamma )}_{m}(\vec{x};n+1),\label{tower}\\
 n&=&1,2,...,p-1, \nonumber
\ea
with energy values $E^{(\gamma )}_m(n+1)$ (see (\ref{energy})).
The restrictions for values of $m$ depend on $\gamma :$
\ba
&&m<(n-A-B)/2,\,\gamma =1;\quad\quad m<(n-1+A-B)/2,\,\gamma=2;\nonumber\\
&& m<(n-1-A+B)/2,\,\gamma=3;\quad m<(n-2+A+B),\, \gamma=4. \nonumber
\ea
This is an appropriate point to remark that the situation of general position corresponds to 
the simple spectrum of $H(p+1),$ which consists of levels $E^{(\gamma )}_k(p+1)$  
(see (\ref{energy}) - (\ref{energy4})) for $\Psi^{(\gamma )}_k(p+1)$ and levels
$E^{(\gamma )}_m(n+1)$ for the states (\ref{tower}). But nothing prohibits from the 
possible occasional degeneracy of the spectrum for some specific values of parameters. 
Indeed, this situation is nongeneric: an occasional degeneracy of levels may occur only for some 
single values of parameters $A, B.$ In such a case, the degeneracy can be removed easily by an arbitrary 
small variations of $A, B.$
%This is an appropriate point to remark that in general, for some specific values of parameters, nothing
%prohibits from the possible degeneracy of the spectrum of $H(p+1),$ which consists of levels $E^{(\gamma )}_k(p+1)$
%(see (\ref{energy}) - (\ref{energy4}))
%for $\Psi^{(\gamma )}_k(p+1)$ and levels $E^{(\gamma )}_m(n+1)$ for the states (\ref{tower}).
%Indeed, the occasional degeneracy of levels is possible for some specific values of parameters $A, B$ and integer $k, m, n, p,$
%but the degeneracy can be avoided easily by choosing suitable values of $A, B.$

The normalizability of functions of the second class is obvious by construction, but this property
for the wave functions (\ref{tower}) will be proven now.
The Hamiltonian $\widetilde{H}(p)$ has the symmetry operator
$\widetilde{R}(p) = Q^-(p)Q^+(p)$ (see (\ref{Rop})), and in turn, $H(p+1) -$
its own symmetry operator $R(p+1) = Q^+(p+1)Q^-(p+1).$ As far as these Hamiltonians coincide
(shape invariance) $H(p+1) =
\widetilde{H}(p),$ the corresponding symmetry operators must
coincide as well, but up to the function of the Hamiltonian. Indeed, by straightforward
calculation one obtains the relation:
\begin{equation}
R(p+1) - \widetilde{R}(p) = 8(2p+1)\biggl( \widetilde{H}(p) +
2(2p^2+2p+1) \biggr), \label{diff}
\end{equation}
which will help to analyze the normalizability.

The norm of the arbitrary wave function $\Psi^{(\gamma )}_{m,(p-n)}(\vec x; p+1)$
(\ref{tower}) can be written as:
\ba
&& \|\Psi^{(\gamma )}_{m,(p-n)}(\vec{x};p+1)\|^2=
\langle \Psi^{(\gamma )}_{m}(\vec{x};n+1)| Q^+(n+1)Q^+(n+2)...Q^+(p)\cdot \nonumber\\
&&\cdot Q^-(p)Q^-(p-1)...Q^-(n+2)Q^-(n+1) \Psi^{(\gamma )}_{m}(\vec{x};n+1)\rangle.
\nonumber
\ea
To simplify it, one may explore Eq.(\ref{diff}) and its consequence:
\ba
&&\biggl( Q^+(n+1)Q^-(n+1) - Q^-(n)Q^+(n) \biggr) Q^-(n)Q^-(n-1) \dots
Q^-(m) =\nonumber\\&&= Q^-(n)Q^-(n-1) \dots Q^-(m) \Gamma_{mn},
\nonumber
\ea
where $\Gamma_{mn}$ is the function of the Hamiltonian:
$$\Gamma_{mn}= 8(2n+1)\biggl( H(m) + 2(2n^2+2n+1) \biggr) .$$
The following relation can be derived by induction:
\ba
 &&Q^+(n+1)Q^+(n+2)...Q^+(p)Q^-(p)Q^-(p-1)...Q^-(n+1)=\nonumber\\
 &&=R(n+1)\biggl(R(n+1)+\Gamma_{n+1,n+1}\biggr)\cdot \biggl(R(n+1)+\Gamma_{n+1,n+1}+\Gamma_{n+1,n+2}\biggr)...\nonumber\\
 &&... \biggl(R(n+1)+\Gamma_{n+1,n+1}+\Gamma_{n+1,n+2}+...+\Gamma_{n+1,p-1}\biggr),
\nonumber
\ea
and finally, one obtains that norms of
the wave functions $\Psi^{(\gamma )}_{m,(p-n)}(\vec x; p+1)$ of second class for $H(p+1)$ are proportional to
the norms of wave functions $\Psi^{(\gamma )}_{m}(\vec x; n+1)$ of the first class for the
Hamiltonian $H(n+1):$
\ba
 &&\|\Psi^{(\gamma )}_{m,(p-n)}(\vec{x};p+1)\|^2=64(2n+1)\biggl( E^{(\gamma )}_m(n+1)+
 2n^2+2n+1\biggr) \cdot\nonumber\\
 &&\cdot\prod_{q=n+1}^{p-1}
 \biggl( (q+1)^2-n^2\biggr) \biggl( E^{(\gamma )}_m(n+1)+2((q+1)^2+n^2)\biggr)
 \|\Psi^{(\gamma )}_{m}(\vec{x};n+1)\|^2.
\nonumber
\ea
It is easy to check explicitly that the
coefficient of proportionality is positive. Hence, as far as the initial state
$\Psi^{(\gamma )}_m(n+1)$ is normalizable by the construction, any wave
function $\Psi^{(\gamma )}_{m,(p-n)}(p+1)$ is normalizable too.

One more statement is necessary to prove in order to be sure that the full variety
of eigenfunctions for $H(p+1)$ was constructed above.
Namely, we must prove that no additional normalizable wave functions exist besides
those in (\ref{D37}), (\ref{D38}). Starting from the lowest Hamiltonians, let us suppose that $H(2)$ has such
additional eigenfunction $\Phi (2)$, which differs from the linear combination of zero modes of $Q^+(1).$
Then, it follows from the intertwining relations, that $Q^+(1)\Phi (2)$ must satisfy the Schr\"odinger equation
with Hamiltonian $H(1).$ One may check that supercharges $Q^{\pm}(p)$ do not change the normalizability neither
at infinity, nor at coordinate axes $x_1, x_2:$ the detail analysis is presented below in the next paragraphs of this Section.
As we already know from Section 3, the Hamiltonian $H(1)$ has no bound states at all, and therefore, our supposition was wrong.
Let us suppose now that the first Hamiltonian possessing
such additional state $\Phi (p+1)$ is $H(p+1),$ while all previous Hamiltonians
$H(p),\,H(p-1),\, ... H(2)$ have bound states of the forms (\ref{D37}), (\ref{D38}), only.
Then, due to intertwining relations, $Q^+(p)\Phi (p+1)$ is the eigenfunction $\Psi (p)$ of $H(p),$
and therefore, coincides either with $\Psi_k(p)$ or with $\Psi_{l,1}(p),$ by our assumption.
For simplicity, we do not consider here the case of possible degeneracy of levels of $H(p)$
(the conclusion will be the same in this case).
Acting by $Q^+(p)$ onto $\Phi (p+1)$ and using the relation (\ref{diff}),
one obtains by straightforward calculations that for both options, $\Psi (p)$ is proportional
to $Q^+\Psi_{n,1}(p+1)$ with some suitable $n.$ Therefore, the wave function $\Phi (p+1)$ coincides
with $\Psi_{n,1}(p+1)$ up to zero modes of $Q^+(p):$
$$\Phi (\vec x; p+1) = c_1(p+1)\Psi_{n,1}(\vec x; p+1) +  c_2(p+1)\Psi_{k}(\vec x; p+1),$$
where $c_{1,2}(p+1)$ are constants.

Thus, the problem is reduced to the question: whether the operators $Q^{\pm}(p)$
are able to change the normalizability of functions. If they are not, no additional
normalizable eigenfunctions of $H(p+1)$ exist.
It is evident from the explicit expressions (\ref{kuplus}) of $Q^{\pm}(p)$
and from taking into account the exponential decreasing of $\Psi$ at infinity,
that $Q^{\pm}$ can not violate integrability of $|\Psi|^2$ at $\pm\infty .$

More difficult problem arises in the neighborhood of $x_1\to 0$ and/or $x_2\to 0.$
The part of $Q^{\pm}$ linear in derivatives coincides with the operator $\hat D,$
defined in (\ref{TT4}). In the limit
$x_2\to 0$, $x_1\neq 0,$ it is:
 $$\hat D \sim 4p(\tanh x_1)\partial_1-4p(1-\tanh^2x_1)x_2\partial_2,$$
i.e. it does not change the asymptotic behaviour of the function.
Analogous conclusion is true in the limit $x_2\to 0$, $x_1\neq 0.$
To analyze the limit when both $x_{1,2}\to 0,$ it is convenient
to use the polar coordinates $x_+=R\cos\varphi ,
x_-=R\sin\varphi .$ Asymptotically, $\hat D$ for $R \to 0$ is:
$$\hat D\sim
 4p(\sin(2\varphi )R\partial_{R}+\cos(2\varphi )\partial_{\varphi }),$$
i.e. in this limit $\hat D$ can not change the behaviour of function as well.

Coming back to the operators $Q^{\pm},$ the only terms which could in principle
change the behaviour of function at $x_1\to 0$ and/or $x_2\to 0$ are:
 $$Q^{\pm}(p)\sim -\biggl( -\partial_1^2-k_2\sinh^{-2}(x_1)\biggr) +
 \biggl( -\partial_2^2 +k_1\sinh^{-2}(x_2)\biggr) .$$
Comparing these terms with (\ref{H}), (\ref{HH}), we observe
the same parts (although with different signs) in asymptotical expressions:
 $$H(p) \sim \biggl( -\partial_1^2- k_2\sinh^{-2}(x_1)\biggr) +
 \biggl( -\partial_2^2 +k_1\sinh^{-2}(x_2)\biggr) .$$
Since $\Psi (\vec x; p)$ are eigenfunctions of $H(p),$
$Q^{\pm}(p)$ are not able to change the behaviour of $\Psi ,$
and the absence of any additional wave functions besides that of (\ref{D37}), (\ref{D38})
types was thus proven.

\section{Examples.}

The explicit expressions for matrix elements $B_{ik}$ from (\ref{B}), which are necessary
to build the eigenfunctions (\ref{D37}) of $H(p+1)$, seem to be difficult to present in a general form.
Nevertheless, the problem can be solved straightforwardly for low values of $p.$
By means of separation of variables, we demonstrated in Section 3, that the Hamiltonian $H(1)$ has no bound states.

The next Hamiltonian $H(2)$ (it corresponds to $p=1$ in formulas above) has two bound states: one bound state with $k^{(1)}=0$, due to
inequality (\ref{condition}) with $p=1,$ and the second bound state with $k^{(2)}=0$ or $k^{(3)}=0,$
due to inequalities (\ref{condition2}), (\ref{condition3}), depending on
the positivity of $(A-B)$ or $(B-A).$ Of course, these bound states are of the second class, i.e. are built from the zero modes:
\ba
&&\Psi^{(1)}_0(\vec x; 2) \sim \widetilde\Omega^{(1)}_0(\vec x; 1)=\pm\left( \cosh(x_+) \cosh(x_-)\right)^{-1}
\cdot\nonumber\\&&\cdot \sinh^{A}(|x_1|) \cosh^{B}(|x_1|) \sinh^{B}(|x_2|) \cosh^{A}(|x_2|)
\label{p=1}
\ea
with energy $E_0^{(1)}(2)=-2\biggl((A+B-1)^2+1\biggr)$, and (for $A>B$)
\ba
&&\Psi^{(2)}_0(\vec x; 2) \sim \widetilde\Omega^{(2)}_0(\vec x; 1)=\pm\left( \cosh(x_+) \cosh(x_-)\right)^{-1}
\cdot\nonumber\\&&\cdot \sinh^{1-A}(|x_1|) \cosh^{B}(|x_1|) \sinh^{B}(|x_2|) \cosh^{1-A}(|x_2|)
\label{pp=1}
\ea
with energy $E_0^{(2)}(2)=-2\biggl((B-A)^2+1\biggr)$. No bound states of the first class exist in this case.

For the next value $p=2,$ i.e. for the Hamiltonian $H(3),$ four bound states are of the second class being built by the zero modes
$$
\Psi^{(\gamma )}_0(\vec x; 3) \sim \widetilde\Omega^{(\gamma )}_0(\vec x; 2); \,\,\gamma =1,2,3,4
$$
with energies
$$
E_0^{(1)}(3)=-2\biggl((A+B-2)^2+1\biggr);\, E_0^{(2)}(3)=-2\biggl((B-A-1)^2+1\biggr);
$$
$$
E_0^{(3)}(3)=-2\biggl((A-B-1)^2+1\biggr);\, E_0^{(4)}(3)=-2\biggl((A+B)^2+1\biggr).
$$
But in this case, two wave functions of the first class also
can be built from (\ref{p=1}) and (\ref{pp=1}) by the procedure (\ref{D38}):
\begin{equation}\label{p=33}
  \Psi^{(\gamma )}_{01}(\vec x; 3)\sim Q^-(2)\Psi^{(\gamma )}_0(\vec x;2);\quad \gamma =1,2 \quad A>B.
\end{equation}
Their energies $E^{(\gamma )}_{01}(3)$ coincide with $E^{(\gamma )}_0(2);\, \gamma =1,2$ above.

The Hamiltonian $H(p+1)$ with $p=3$ for $A>B$ has six states which are built from bound states of $H(3)$ by means of operator $Q^-(4).$
They have the same energies $E_0^{(1)}(3), E_0^{(2)}(3), E_0^{(3)}(3), E_0^{(4)}(3)$ and $E_{0}^{(1)}(2), E_0^{(2)}(2).$
As for the second class bound states, six such bound states exist: $k^{(1)}=0, 1;\, k^{(2)}=0, 1;\, k^{(3)}=k^{(4)}=0.$
This set includes two wave functions coinciding with $\widetilde\Omega^{(3)}_0(\vec x;4);\,\widetilde\Omega^{(4)}_0(\vec x;4),$ and four other
wave functions have to be built as linear combinations of pairs of zero modes $\widetilde\Omega^{(1)}_0(\vec x;4),\,\widetilde\Omega^{(1)}_1(\vec x;4)$ and $\widetilde\Omega^{(2)}_0(\vec x;4),\,\widetilde\Omega^{(2)}_1(\vec x;4),$
since $N^{(1)}=N^{(2)}=1$ for $p=4.$ The matrix elements of $2\times 2$ triangular matrix $C^{(1)}_{ki}$ are defined by (\ref{DD16}), (\ref{DD17}):
$$C^{(1)}_{00}=E^{(1)}_0(4)=-2\biggl[(A+B-3)^2+9\biggr]; \quad C^{(1)}_{11}=E^{(1)}_1(4)=-2\biggl[(A+B-1)^2+9\biggr];\quad
C^{(1)}_{10}=-24.$$
These elements are necessary to determine coefficients $B^{(1)}_{ki}$ for (\ref{D37}). From Eq.(\ref{B}),
one can find that for $C^{(1)}_{00}\neq C^{(1)}_{11},$ as in the present case, the matrix $B^{(1)}_{ki}$ is also triangular $B^{(1)}_{01}=0$.
More of that, while
$$B^{(1)}_{10}=\frac{C^{(1)}_{10}}{C^{(1)}_{11}-C^{(1)}_{00}}B^{(1)}_{11},$$
the last coefficient $B^{(1)}_{00}$ is arbitrary. This fact is not discouraging, since the linear combination (\ref{D37})
for $\Psi^{(1)}_0(4)$ includes only one term (with arbitrary $B^{(1)}_{00}$). Thus, the value of $B^{(1)}_{00}$ is fixed by
unity norm of $\Psi^{(1)}_0(4).$ The second linear combination (\ref{D37}) for $\Psi^{(1)}_1(4)$ includes both $B^{(1)}_{10}$
and $B^{(1)}_{11},$ which are proportional to each other. The absolute values of these coefficients will also be fixed
by normalization of the wave function. Analogous calculations can be easily repeated for $\gamma = 2.$

\section{Conclusions.}

An exhaustive procedure of analytical solution of two-dimensional generalization of P\"oschl-Teller model with integer values
of parameter $p$ was presented above. Being based on SUSY intertwining relations and shape invariance of the model, the procedure
replaces the standard method of separation of variables which is not applicable here, and it can be considered as a special - SUSY - separation of variables.

In order to confirm obtained results for $H(p+1)$, it is useful to compare them with the limiting case which possesses the direct solution by means
of separation of variables. Indeed, if the parameters $A, B,$ which originally belong to the interval $(0,\, 1/2),$ are chosen on the limit of range $A,B \to 0,$ the procedure above (starting from Sect. 3) does not work. But due to conventional separation of variables,
%When the parameters $A, B,$ which belong to the interval $(0,\, 1/2),$ are chosen on the limit of range
%$A,B \to 0,$
the Hamiltonian $H(p+1)$ from (\ref{H}) is reduced (up to a trivial multiplier $2$) to a sum of two one-dimensional Hamiltonians with
well known reflectionless potentials in variables $x_{\pm}:$
$$
h(p+1)(x)=-\partial^2 - \frac{p(p+1)}{\cosh^2x}.
$$
The spectra of these Hamiltonians are well known: for $p \in [L, L+1)$ they have exactly $L$ bound states. To compare the properties of spectra with that in Section 7, one has to explore the original inequalities (\ref{condition}), (\ref{condition2}) - (\ref{condition4}) and (\ref{energy}),
(\ref{energy2}) - (\ref{energy4}) where $A$ and $B$ are written explicitly. The point is that some of bound states described in Section 7 disappear
in the limit $A, B \to 0.$

Thus, $H(p+1)$ with $p=1$ has one bound state with energy $E_0^{(1)}(2)=-4$ in the limiting case, since only (\ref{condition}) can be satisfied with $k^{(1)}=0.$ Taking into account the multiplier $2$ mentioned above, this value of energy coincides with double value of eigenvalue
of $h(2).$

For $p=2,$ three bound states of the second class for $H(p+1)$ exist in the limiting case, with $k^{(1)}=k^{(2)}=k^{(3)}=0$ and energies
$E^{(1)}_0(3)=-16;\, E^{(2)}_0(3)=E^{(3)}_0(3)=-10,$ and one bound state of the first class with energy $E^{(1)}_0(2)=-4.$
The one-dimensional Hamiltonian $h(3)$ has two bound states, leading just to four possible bilinear combinations in the standard separation of variables.

For $p=3$ with $A>B$ the limiting Hamiltonian $H(4)$ has nine bound states: five of second class $k^{(1)}=0, 1; k^{(2)}=k^{(3)}=k^{(4)}=0,$ and additionally four bound states of first class inherited from four bound states of $H(3).$ As it should be, this number coincides with $3 \times 3 = 9$ possible combinations of one-dimensional wave functions.

%Finally, it is necessary to note that the results of the paper might have twofold destination. First,
%to extend the pure analytical methods of analysis of two-dimensional Quantum Mechanics,
%which are very restrictive. Second, to use them in future for quantum design in different applications,
%such as quantum dots, modern nanodevices and some cosmological models.
Finally, it is necessary to note that the results of the paper might have several destinations. First, 
to realize new pure analytical methods of analysis of two-dimensional Quantum Mechanics, which are few 
in number to present day. Second, to use these new methods for quantum design in different applications, 
such as quantum dots, modern nanodevices and some cosmological models. Third, successful use of one-dimensional 
P\"oschl-Teller potential for description of interaction in diatomic molecules signals about perspectives 
to use its two-dimensional analogues in quantum chemistry. Fourth, the complete solvability of the present 
model gives the opportunity to check the validity of different approximate schemes in many-particle quantum physics.

\section*{Acknowledgements.}

The work was partially supported by the RFFI grant 09-01-00145-a (M.V.I. and
P.A.V.). P.A.V. is indebted to the International Centre of Fundamental Physics in
Moscow and the non-profit foundation "Dynasty" for financial support.

\section*{Appendix: Calculation of the Coefficients $C_{ki}$.}

We will calculate the coefficients $C^{(1)}_{ki},$ but $C^{(\gamma )}_{ki}$ with $\gamma = 2,3,4$ can be easily obtained by replacing in formulas below $A \to (1-A)$ etc. The calculation will be started from Eq.(\ref{6}), where parameters are defined according
to (\ref{def}), (\ref{ak}), (\ref{666}).
It is useful to express the hypergeometric functions (with $a_k=-k$)
in terms of Jacobi polynomials:
\ba
 &&_2F_1(a_k,b_k;c_1;z_1)=\frac{k!}{(\alpha
 +1)_k}P^{(\alpha,\beta)}_k(y_1),\nonumber\\&&
 _2F_1(a_k+1,b_k+1;c_1+1;z_1)=\frac{(k-1)!}{(\alpha +2)_k}
 P^{(\alpha+1,\beta+1)}_{k-1}(y_1),
\nonumber\\
 &&_2F_1(a_k,b_k;c_2;z_2)=\frac{k!}{(\beta
 +1)_k}P^{(\beta,\alpha)}_k(y_2),\nonumber\\&&
 _2F_1(a_k+1,b_k+1;c_2+1;z_2)=\frac{(k-1)!}{(\beta +2)_k}
 P^{(\beta+1,\alpha+1)}_{k-1}(y_2),
\nonumber
\ea
where Pokhgammer symbols $(\Gamma)_k$ are defined as $(\Gamma)_k\equiv \Gamma (\Gamma+1)\cdot ... (\Gamma+k-1),$
and new variables and new parameters will be more suitable:
 $$y_{1,2}\equiv 1-2z_{1,2},\qquad \alpha \equiv A-\frac{1}{2},\quad \beta
 \equiv B-\frac{1}{2},$$
and now:
 $$b_k=1+\alpha+\beta +k,\quad c_1=1+\alpha, \quad
 c_2=1+\beta,\quad \sqrt{-e_k}=1+\alpha+\beta+2k.$$
Taking into account, that $(\alpha+1)(\alpha+2)_{k-1}=(\alpha+1)_k,$
the l.h.s. of (\ref{6}) becomes:
\ba
M\equiv\frac{(k!)^2}{(\alpha+1)_k(\beta+1)_k}\biggl\{2\biggl( 2p(1+\alpha+\beta)-
 (1+\alpha+\beta+2k)^2-2p^2\biggr) P^{(\alpha,\beta)}_k(y_1)\,P^{(\beta,\alpha)}_k(y_2)-\nonumber\\
 -\frac{4p(1+\alpha+\beta+k)}{y_1+y_2}\biggl( (1-y_1^2)
 P^{(\alpha+1,\beta+1)}_{k-1}(y_1)\,P^{(\beta,\alpha)}_k(y_2) + (1-y_2^2)\,
 P^{(\beta+1,\alpha+1)}_{k-1}(y_2)\,P^{(\alpha,\beta)}_k(y_1)\biggr) \biggr\}.\label{DD6}
\ea

Let us use the relation for Jacobi polynomials 22.17.15 from \cite{abramovich}
with $n$ replaced by $n+1$ and $\beta$ by $\beta -1.$ Multiplying it by $(1+x)$
and using the relation 22.7.16 from \cite{abramovich}, one obtains:
 \ba
 \biggl( n+\frac{\alpha+\beta+1}{2}\biggr) (1-x^2) P^{(\alpha+1,\beta+1)}_{n-1}(x) =
 \frac{2(n+\alpha)(n+\beta)}{2n+\alpha+\beta}P^{(\alpha,\beta)}_{n-1}(x)-\nonumber\\
 -\frac{2n(n+1)}{2(n+1)+\alpha+\beta}P^{(\alpha,\beta)}_{n+1}(x)+
 2\biggr(\frac{2n(n+\alpha)}{2n+\alpha+\beta}-\frac{n(n+1+\beta)}{2(n+1)+
 \alpha+\beta}\biggl)P^{(\alpha,\beta)}_n(x).\label{DD7}
 \ea
Let us write the same relation but for $x\Longrightarrow y$ and
$\alpha \Longleftrightarrow \beta ,$ and add it to the initial Eq.(\ref{DD7}).
Then, rewriting $P_{n+1}$ as a combination of $P_{n-1}$ and $P_n$
(according to 22.7.1 from \cite{abramovich}), we obtain:
\ba
 &&(1-x^2)
 P^{(\alpha+1,\beta+1)}_{n-1}(x)P^{(\beta,\alpha)}_n(y)+(1-y^2)
 P^{(\beta+1,\alpha+1)}_{n-1}(y)P^{(\alpha,\beta)}_n(x)=\nonumber\\
 &&=\frac{2}{n+1+\alpha+\beta}\biggr\{\frac{2(n+\alpha)(n+\beta)}{2n+\alpha+\beta}
 \biggr(P^{(\alpha,\beta)}_{n-1}(x)P^{(\beta,\alpha)}_n(y)+
 P^{(\beta,\alpha)}_{n-1}(y)P^{(\alpha,\beta)}_n(x)\biggl)-\nonumber\\
 &&-n(x+y)P^{(\alpha,\beta)}_n(x)P^{(\beta,\alpha)}_n(y).\biggl\}.
\nonumber
 \ea
Due to this relation, (\ref{DD6}) takes the form:
 \ba
 M=\frac{(k!)^2}{(\alpha+1)_k(\beta+1)_k}\biggr\{-2\biggr((1+\alpha+\beta+2k-p)^2
 +p^2\biggl)P^{(\alpha,\beta)}_k(y_1)P^{(\beta,\alpha)}_k(y_2)-\nonumber\\
 -\frac{16p(k+\alpha)(k+\beta)}{2k+\alpha+\beta}\frac{1}{y_1+y_2}
 \biggr(P^{(\alpha,\beta)}_{k-1}(y_1)P^{(\beta,\alpha)}_k(y_2)+
 P^{(\beta,\alpha)}_{k-1}(y_2)P^{(\alpha,\beta)}_k(y_1)\biggl)
 \biggl\}.\label{DD9}
 \ea
The r.h.s. of (\ref{DD9}) includes the combination:
 $$\Phi_{k-1,k}\equiv P^{(\alpha,\beta)}_{k-1}(y_1)P^{(\beta,\alpha)}_k(y_2)+
 P^{(\beta,\alpha)}_{k-1}(y_2)P^{(\alpha,\beta)}_k(y_1),$$
 which (by means of recurrent formula 22.7.1 from \cite{abramovich}) satisfy:
 \ba
 \Phi_{k-1,k}=(y_1+y_2)\frac{a_{3(k-1)}}{a_{1(k-1)}}P^{(\alpha,\beta)}_{k-1}(y_1)
 P^{(\beta,\alpha)}_{k-1}(y_2)-\frac{a_{4(k-1)}}{a_{1(k-1)}}\Phi_{k-1,k-2},\label{DD11}
 \ea
 where the following definitions were introduced:
 \ba
&&a_{1n}=2(n+1)(n+\alpha+\beta+1)(2n+\alpha+\beta);\nonumber\\
&&a_{2n}=(2n+\alpha+\beta+1)(\alpha^2-\beta^2);\nonumber\\
&&a_{3n}=(2n+\alpha+\beta )(2n+\alpha+\beta +1)(2n+\alpha+\beta +2);\nonumber\\
&&a_{4n}=2(n+\alpha)(n+\beta)(2n+\alpha+\beta+2).\nonumber
\ea
Since
\be
 \Phi_{0,1}=(y_1+y_2)\frac{a_{30}}{a_{10}}P^{(\alpha,\beta)}_0(y_1)
 P^{(\beta,\alpha)}_0(y_2),\label{DD12}
\ee
Eqs.(\ref{DD11}), (\ref{DD12}) give:
 \ba
 \Phi_{k-1,k}=(y_1+y_2)\biggr(\frac{a_{3(k-1)}}{a_{1(k-1)}}P^{(\alpha,\beta)}_{k-1}(y_1)
 P^{(\beta,\alpha)}_{k-1}(y_2)-\frac{a_{3(k-2)}}{a_{1(k-2)}}\frac{a_{4(k-1)}}{a_{1(k-1)}}
 P^{(\alpha,\beta)}_{k-2}(y_1)P^{(\beta,\alpha)}_{k-2}(y_2)+\nonumber\\
 +\frac{a_{3(k-3)}}{a_{1(k-3)}}\frac{a_{4(k-1)}}{a_{1(k-1)}}\frac{a_{4(k-2)}}{a_{1(k-2)}}
 P^{(\alpha,\beta)}_{k-3}(y_1)P^{(\beta,\alpha)}_{k-3}(y_2)+...(-1)^{k-1}
 \frac{a_{30}}{a_{10}}\frac{a_{4(k-1)}...a_{41}}{a_{1(k-1)}...a_{11}}P^{(\alpha,\beta)}_0(y_1)
 P^{(\beta,\alpha)}_0(y_2)\biggl).\label{DD13}
 \ea
The coefficients
 $$b_{ki}\equiv (-1)^{k+i-1}\frac{a_{3i}}{a_{1i}}\frac{a_{4(k-1)}a_{4(k-2)}...a_{4(i+1)}}
 {a_{1(k-1)}a_{1(k-2)}...a_{1(i+1)}},\quad b_{k,k-1}\equiv
 \frac{a_{3(k-1)}}{a_{1(k-1)}}$$
allow to write (\ref{DD13}) more compactly:
 $$\Phi_{k-1,k}=(y_1+y_2)\sum_{i=0}^{k-1}b_{ki}P^{(\alpha,\beta)}_{k}(y_1)
 P^{(\beta,\alpha)}_{k}(y_2).$$
Finally, substituting it into (\ref{DD6}), we obtain from (\ref{6})
the general expressions for desired coefficients $C^{(1)}_{ki}:$
 \ba
 C^{(1)}_{kk}=-2\biggr((1+\alpha+\beta+2k-p)^2+p^2\biggl),\label{DD16}\\
 C^{(1)}_{k,i<k}=-\frac{16(k!)^2p(k+\alpha)(k+\beta)(1+\alpha)_i(1+\beta)_i}
 {(i!)^2(1+\alpha)_k(1+\beta)_k(2k+\alpha+\beta)}b_{ki}.\label{DD17}
 \ea

\end{document}